\documentclass[aps,prl,twocolumn,superscriptaddress,groupedaddress]{revtex4}

\usepackage{amsmath,amssymb,bm,color,comment,dcolumn,mathrsfs,tabularx,graphicx}

\newcommand{\al}[1]{\ifdefined\compile \begin{align} #1 \end{align}\fi}
\def\l{\ell}

\def\ol{\overline}
\def\hatn{\Omega}
\def\bD{\ol{d}}
\def\uD{\ol{D}}
\def\hD{\widehat{d}}
\def\Dmin#1{D_{#1}^{\rm min}}
\def\Dmax#1{D_{#1}^{\rm max}}
\def\dg{\delta}
\def\mag{\kappa}
\def\hmag{\widehat{s}}
\def\nmag{\epsilon}
\def\pdf{p}
\def\bx{\bm{x}}
\def\ux{\ol{\bm{x}}}
\def\ave#1{\langle #1 \rangle}
\newcommand{\INT}[5]{%
\ifx#3_%
{\int_{#4}^{#5} \!\! {\rm d}^{#1} {#2} \,}
\else
{\int_{#4}^{#5} \!\! \frac{{\rm d}^{#1} {#2}}{#3} \,} 
\fi %
} 
\newcommand{\bc}{\begin{center}}
\newcommand{\ec}{\end{center}}
\newcommand{\bi}{\begin{itemize}}
\newcommand{\ei}{\end{itemize}}
\newcommand{\mC}[1]{\mathcal{#1}}
\newcommand{\compile}{}

\begin{document}

\ifdefined\compile %compile

% Title %
\title{Anisotropies of Gravitational-Wave Standard Sirens as a New Cosmological Probe without Redshift Information}

% Authors %
\author{Toshiya Namikawa}
\affiliation{Department of Physics, Stanford University, Stanford, California 94305, USA}
\affiliation{Kavli Institute for Particle Astrophysics and Cosmology, SLAC National Accelerator
Laboratory, Menlo Park, California 94025, USA}
\author{Atsushi Nishizawa}
\affiliation{Theoretical Astrophysics 350-17, California Institute of Technology, Pasadena, California 91125, USA}
\author{Atsushi Taruya}
\affiliation{Yukawa Institute for Theoretical Physics, Kyoto University, Kyoto 606-8502, Japan}
\affiliation{Kavli Institute for the Physics and Mathematics of the Universe,
Todai Institutes for Advanced Study, the University of Tokyo, Kashiwa, Chiba 277-8583, Japan}

% Date %
\date{\today}

% Preprint %
%\preprint{YITP-15-XX}

% Abstract %
\begin{abstract}
Gravitational waves (GWs) from compact binary stars at cosmological distances are promising and 
powerful cosmological probes, referred to as the GW standard sirens. With future GW detectors, 
we will be able to precisely measure source luminosity distances out to a redshift $z\sim5$. 
To extract cosmological information, previously proposed cosmological studies 
using the GW standard sirens rely on source redshift information obtained through an extensive 
electromagnetic follow-up campaign. However, the redshift identification is typically time consuming 
and rather challenging. Here, we propose a novel method for cosmology with the GW standard sirens 
free from the redshift measurements. Utilizing the anisotropies of the number density and luminosity 
distances of compact binaries originated from the large-scale structure, we show that,
once GW observations will be well established in the future, 
(i) these anisotropies can be measured even at very high redshifts ($z\geq 2$), 
where the identification of the electromagnetic counterpart is difficult, 
(ii) the expected constraints on the primordial non-Gaussianity with the Einstein Telescope would be 
comparable to or even better than the other large-scale structure probes at the same epoch, and 
(iii) the cross-correlation with other cosmological observations is found to have high-statistical 
significance, providing additional cosmological information at very high redshifts. 
\end{abstract}

% PACS %
%\pacs{}

% Output Title/Abstract etc %
\maketitle

\fi % compile

% Contents %

%////////////////////////////////////////////////////////////////////////////////////////////////////%
{\it Introduction.---}
% Status and cosmological application of GW observations
%The second-generation gravitational wave (GW) detectors such as aLIGO, aVIRGO, and KAGRA will start 
%operations within one or two years, and will achieve the first direct detection of 
%a GW \cite{Evans2014GRG}. After successful detections, the GW would become a powerful 
%and promising tool to probe cosmology and astrophysics, complementary to or even independent of 
%the electromagnetic observations. 
First detection of gravitational waves (GW) by the advanced laser interferometer, aLIGO, has opened 
a new window to probe unseen Universe \cite{GW150914}. With a network of detectors including 
aVIRGO and KAGRA \cite{Evans2014GRG}, the GW would become a powerful and promising tool to probe 
cosmology and astrophysics, complementary to or even independent of the electromagnetic observations. 
Indeed, the future ground- and space-based GW experiments such as the Einstein Telescope (ET) 
\cite{Punturo:2010}, 40-km LIGO \cite{Dwyer2015PRD}, eLISA \cite{Seoane:2013qna}, and DECIGO 
\cite{Sato:2009}, are planning to realize much higher sensitivity, with which we can observe a large 
number of neutron star (NS) binaries at cosmological distances. One important aspect of future GW 
observations is that one would be able to measure the luminosity distance to each binary source 
(so-called {\it standard sirens}) with an unprecedented precision, from which the cosmic expansion 
history can be accurately determined, e.g., Refs.~
\cite{Schutz1986Nature,Petiteau2011ApJ,Cutler:2009qv,Nishizawa:2010xx,Sathyaprakash:2009xt,Taylor:2012db}. 
In addition, measurements of these GWs will be very useful to study the gravitational lensing effect 
induced by the Large-Scale Structure (LSS) by looking at the anisotropies of the observed luminosity distances 
\cite{Cutler:2009qv,Hirata:2010,Shang2011MNRAS,Camera:2013xfa}. 
The use of the LSS-induced anisotropies on other cosmological probes has been also studied
in Refs.~\cite{Cooray:2006,McQuinn:2013} in different context.

Nevertheless, one crucial assumption behind these studies is that redshift information or a 
corresponding distance measure other than luminosity distance are {\it a priori} known because GW 
observation alone is not sensitive to source redshift. However, the source identification and 
redshift measurement with electromagnetic (EM) follow-up observations are rather challenging, and 
extensive follow-up campaigns are required. In particular, the EM observation to identify the host 
galaxy of each compact binary would be very difficult at higher redshifts \cite{Nissanke:2013}, and 
the feasibility of source identification largely depends on the emission mechanism of the EM counterparts. 
As discussed in \cite{Nishizawa:2011eq}, the success rate of identification is estimated 
to be $\sim10^{-4}$ from galaxy catalogs of future surveys and $\sim 10^{-3}$ from coincident searches of 
short gamma-ray bursts with gamma-ray telescopes. 
To circumvent the situations, alternative methods have been proposed. 
Reference \cite{MacLeod2008PRD} presents a statistical method without identifying an EM counterpart, 
assuming a redshift distribution based on a complete galaxy catalog. 
References \cite{Messenger:2011gi,Taylor:2011fs} assume the equation of state of neutron stars and/or 
a narrow distribution of NS mass to infer the redshift of each source. 
However, the reliability of redshift estimation largely depends on the underlying assumptions. 

% Our idea
In this {\it Letter}, we propose a novel approach to pursue the cosmology with the GW standard sirens 
without any assumptions for redshift information. The proposed method is to utilize anisotropies in 
the observables induced by the LSS. The distribution function is anisotropic due to the clustering of 
the NS binaries by the fluctuations of the gravitational potential. The LSS also induces the 
gravitational lensing, and the measured luminosity distance is modified. These anisotropic signals 
contain rich cosmological information helpful to constrain the cosmic expansion history and structure 
formation. One remarkable difference from the previous studies is that the present method directly 
offers a redshift-free measurement of the anisotropies at high statistical significance. The measured 
signal is expected to be powerful to constrain cosmology especially at the distant universe. In the 
followings, we discuss how the LSS induces the anisotropies in the observables constructed from GW signals.

%////////////////////////////////////////////////////////////////////////////////////////////////////%
{\it Anisotropies induced by the LSS.---}
An observable considered in this {\it Letter} is a distribution of NS binaries per luminosity distance
$N(D,\hatn)$ as a function of their luminosity distance $D$ and direction $\hatn$. Hereafter, we 
denote the arguments by $\bx=(D,\hatn)$. We then define a normalized distribution function, 
$\pdf(\bx)\equiv N(\bx)/\int {\rm d}D N(\bx)$. At each angular pixel, observed binaries are divided 
into subsamples according to their luminosity distance. We denote, respectively, by $\Dmax{i}$ and 
$\Dmin{i}$ the maximum and minimum values of the observed luminosity distance in $i$th luminosity 
distance bin. 

There are two types of LSS-induced anisotropies in the observables. The clustering of the NS binaries 
at each direction causes the fluctuations in $N(\bx)$. The observed luminosity distance $D$ should 
also have explicit directional dependence by the gravitational lensing of the LSS, and is related to 
the {\it original} luminosity distance $\ol{D}$ (in the absence of the lensing effect) through 
(e.g., Refs.~\cite{FutamaseSasaki:1989,Hirata:2010})
%----------------------------------------------------------------------------------------------------%
\al{ 
	D(\bx) = \uD[1+\mag(\ux)] \;, \quad (\ux=(\uD,\hatn))  \,. \label{mag} 
}
%----------------------------------------------------------------------------------------------------%
The quantity $\mag(\bx)$ is the lensing convergence induced by the gravitational potential of the LSS. 
The lensing effect on the trajectory of GW propagation is at second order in $\mag$, 
and is ignored in the following derivation. 

To discuss the feasibility of measuring the anisotropies from the LSS, we construct an estimator based 
on the observed luminosity distance. First, we may take the average of $D$ within 
the $i$th luminosity-distance bin at each angular direction $\hatn$: 
%----------------------------------------------------------------------------------------------------%
\al{
	\hD_i(\hatn) &= \INT{}{D}{_}{\Dmin{i}}{\Dmax{i}} D\, \pdf(\bx)  \,. \label{Eq:ObsD}
}
%----------------------------------------------------------------------------------------------------%
Further averaging $\hD_i(\hatn)$ over the entire sky, we obtain a mean luminosity distance $\hD_i$ 
at the $i$th bin. We then estimate the fluctuation in the $i$th luminosity-distance bin at each direction 
by taking the difference $\hD_i(\hatn)-\hD_i$. 
Thus, a simple dimensionless estimator for the LSS-induced anisotropies $\hmag$ is introduced: 
%----------------------------------------------------------------------------------------------------%
\al{
	\hmag_i(\hatn) &\equiv \frac{\hD_i(\hatn)-\hD_i}{\hD_i}  \,. \label{Eq:est-mag}
} 
%----------------------------------------------------------------------------------------------------%
This quantity is proportional to the number density of the NS binaries, and also probes 
the lensing modification to the luminosity distance. Therefore, the above quantity is one of 
the estimators to probe the LSS anisotropies of both the clustering and lensing. 

To understand how the estimator $\hmag_i$ is sensitive to the cosmology, we rewrite Eq.~\eqref{Eq:est-mag} 
in terms of the fluctuations of the number density $\delta$ and lensing convergence field $\mag$ 
generated by the LSS. Here and in what follows, we consider the terms up to the first order of 
$\dg$ and $\mag$. Unlike the methods using the source redshift information, the source distribution 
is given as a function of the observed luminosity distance. Let us recall that the observed number 
distribution is modified by the lensing effect. It is related to the unlensed binary distribution 
$\ol{\pdf}$ through the number conservation 
%----------------------------------------------------------------------------------------------------%
\al{
	\pdf(\bx)\,{\rm d}\Omega\, {\rm d}D = \ol{\pdf}(\uD)[1+\dg(\ux)]\, {\rm d} \Omega\, {\rm d} \uD
	\,. \label{Eq:photon-cons}
}
%----------------------------------------------------------------------------------------------------%
Here the fluctuations of the unlensed quantity $\dg(\ux)$ come from the effect of the NS-binary 
clustering. Introducing a dimensionless quantity $n(\bx)\equiv D\pdf(\bx)$, the above equation implies 
%----------------------------------------------------------------------------------------------------%
\al{
	& n(\bx) = ({\rm d}\uD/{\rm d}D) [1+\mag(\ux)]\ol{n}(\uD)[1+\dg(\ux)] \notag \\
	& \quad \simeq [1-D\mag'(\bx)]\ol{n}[D/(1+\mag(\bx))][1+\dg(\bx)]  \,.
}
%----------------------------------------------------------------------------------------------------%
where the prime denotes the derivative with respect to $D$, 
and we use Eq.~\eqref{mag} and ignore higher-order terms of the fluctuations. 
Substituting the above equation into Eq.~\eqref{Eq:ObsD}, we find
%----------------------------------------------------------------------------------------------------%
\al{ 
	&\hD_i(\hatn) = \INT{}{D}{_}{\Dmin{i}}{\Dmax{i}} 
		\ol{n}\left[\frac{D}{1+\mag(\bx)}\right][1+\dg(\bx)-D\mag'(\bx)]
	\,. \label{Eq:ObsD-mod}
} 
%----------------------------------------------------------------------------------------------------%
Averaging $\hD_i(\hatn)$ over all directions $\hatn$, we obtain a theoretical expression 
for the averaged luminosity distance as
%----------------------------------------------------------------------------------------------------%
\al{
	\hD_i &= \INT{}{D}{_}{\Dmin{i}}{\Dmax{i}} \ol{n}(D) \equiv \bD_i  \,. \label{Eq:est-denom}
}
%----------------------------------------------------------------------------------------------------%
Note that we implicitly assumed that the average of $\mag(\bx)$ over the entire sky becomes zero.
On the other hand, the luminosity-distance anisotropies are obtained by expanding 
Eq.~\eqref{Eq:ObsD-mod} upto first order of $\mag$ and $\dg$ as
%----------------------------------------------------------------------------------------------------%
\al{
	&\hD_i(\hatn) \simeq \bD_i + \INT{}{D}{_}{\Dmin{i}}{\Dmax{i}} \ol{n}(D)[\dg(\bx)+\gamma(D)\mag(\bx)] \,.
}
%----------------------------------------------------------------------------------------------------%
Here we define $\gamma(D) \equiv 1+D[\delta_{\rm D}(D-\Dmin{i})-\delta_{\rm D}(D-\Dmax{i})]$
with $\delta_{\rm D}$ being the delta function. The above equation leads to
%----------------------------------------------------------------------------------------------------%
\al{
	\hmag_i(\hatn) &= \frac{1}{\bD_i}\INT{}{D}{_}{\Dmin{i}}{\Dmax{i}} 
		\ol{n}(D)[\dg(\bx)+\gamma(D)\mag(\bx)]
	\,. \label{Eq:est-theory}
}
%----------------------------------------------------------------------------------------------------%

The anisotropic signal $\hmag_i$ statistically contains rich cosmological information. To extract the 
information, we may move to the harmonic space, and define the angular power spectrum, 
$\widehat{C}^{s_is_j}_\l\equiv\sum_{m=-\l}^\l(\hmag_{i,\l m}\hmag_{j,\l m}^*+{\rm c.c.})/[2(2\l+1)]$, 
where $\hmag_{i,\l m}$ is the spherical harmonic coefficient of the signal. 
Taking the ensemble average of this, the power spectrum is theoretically expressed as
%----------------------------------------------------------------------------------------------------%
\al{
	& C^{s_is_j}_\l \equiv \ave{\widehat{C}^{s_is_j}_\l} =
		\INT{}{D_1}{\bD_i}{\Dmin{i}}{\Dmax{i}} \INT{}{D_2}{\bD_j}{\Dmin{j}}{\Dmax{j}} 
	\notag \\
		&\qquad \times \ol{n}(D_1)\ol{n}(D_2)\bigg\{
		C_\l^{\dg_1\dg_2} + \gamma(D_1)\gamma(D_2) \,C_\l^{\mag_1\mag_2} 
	\notag \\
		&\qquad\qquad + \gamma(D_1)\,C_\l^{\dg_2\mag_1} + (1\leftrightarrow 2) \bigg\}
	\,.
}
%----------------------------------------------------------------------------------------------------%
Here, $C_\l^{\delta\delta}$, $C_\l^{\delta\mag}$, and $C_\l^{\mag\mag}$ are the auto and cross angular 
power spectra of the number density fluctuations and lensing convergence, given by 
$\ave{X_{\l m}(D_1)Y^*_{\l'm'}(D_2)}=\delta_{\l\l'}\delta_{mm'}\,C_\l^{X_1Y_2}$, with $X$ and $Y$ 
being either $\dg$ or $\mag$.
The information on the cosmic expansion and the growth of structure is encapsulated in these power spectra. 
In computing the above quantities theoretically, we need to know the unlensed binary distribution 
$\ol{n}(\uD)$, which is not the actual observable. At first order in $\dg$ and $\mag$, however, 
it can be estimated by averaging the observed distribution over the entire sky.

%////////////////////////////////////////////////////////////////////////////////////////////////////%
{\it Signal-to-noise ratio.---}
Feasibility to measure the power spectrum largely depends on the noise properties of $\hmag_i$. One 
important noise is the measurement error of the luminosity distance coming from the limited 
sensitivity of the GW detector. Let us denote this measurement error for each source by $\nmag=\delta D/D$. 
For simplicity, we assume that $\nmag$ is the random Gaussian field with zero mean, independently of 
the angular position, and it does not correlate between different GW sources. The magnitude of the 
error $\nmag$ does actually depend not only on the detector sensitivity but also on how far each GW 
source is. To estimate the expected size of $\nmag$, we consider the ET, and adopt the sky-averaged 
sensitivity in Ref.~\cite{Regimbau:2012ir}. For each binary source, we assume the restricted $1.5$ 
post-Newtonian waveform, setting the spin parameter to zero. Then, the size of the error $\nmag$ is estimated 
as a function of luminosity distance based on the Fisher matrix analysis presented in Ref.~\cite{Nishizawa:2011eq}. 

Once we obtained the error $\nmag$, we propagate it to the noise of LSS-induced anisotropies as follows. 
Ignoring the LSS effects, the measured luminosity distance is given by $D=\uD(1+\nmag)$. 
This produces an error in the luminosity distance of Eq.~\eqref{Eq:ObsD}, as 
%----------------------------------------------------------------------------------------------------%
\al{
	\delta\hD_i(\hatn) &= \INT{}{D}{_}{\Dmin{i}}{\Dmax{i}} \{\ol{n}(D)-\ol{n}[D/(1+\nmag)]\}
	\notag \\
		&\simeq \INT{}{D}{_}{\Dmin{i}}{\Dmax{i}} \ol{n}'(D) \,D\,\nmag + \mC{O}(\nmag^2)
	\,. \label{Eq:hDi-error}
} 
%----------------------------------------------------------------------------------------------------%
On the other hand, the noise in the mean distance $\hD_i$, obtained by further averaging $\hD_i(\hatn)$ 
over the entire sky, would be of the second order of $\nmag$, and it can be ignored. Thus, from 
Eq.~\eqref{Eq:est-mag}, the noise in $\hmag$ is estimated to be 
$\delta\hmag_i(\hatn)\simeq \delta\hD_i(\hatn)/\bD_i$. 
This produces a shot noiselike contribution, and leads to a systematic offset in the power spectrum, i.e., 
$\ave{\widehat{C}_\l^{s_is_j}}\to C^{s_is_j}_\l +\delta_{ij}\,(\sigma^2_i/N_i)$, 
where $\sigma^2_i$ is defined by
%----------------------------------------------------------------------------------------------------%
\al{
	\sigma_i^2 = \frac{1}{\bD_i^2}\INT{}{D}{_}{\Dmin{i}}{\Dmax{i}} \ol{\pdf}(D)
		\left[\frac{\ol{n}'(D)}{\ol{\pdf}(D)}\right]^2 D^2\sigma^2(D)
	\,. 
}
%----------------------------------------------------------------------------------------------------%
with $\sigma(D)$ being the rms of $\nmag$ estimated based on the Fisher matrix. 
The $N_i$ is the mean number density of GW sources per steradian at $i$th bin. 

Note that the uncertainty coming from the above contribution still remains nonvanishing even if we 
subtract the mean value from the measured power spectrum, and thus needs to be properly taken into 
account in the statistical analysis. Hence, including further the uncertainty coming from the cosmic 
variance, the cumulative signal-to-noise ratio (SNR) for the LSS-induced anisotropies at the $i$th 
bin $\hmag_i$ is defined by 
%----------------------------------------------------------------------------------------------------%
\al{
	\left(\frac{S}{N}\right)^2_{<\l} &\equiv \sum_\l\frac{2\l+1}{2}
		\left[\frac{C^{s_is_i}_\l}{(\sigma^2_i/N_i)+C^{s_is_i}_\l}\right]^2
	\,. \label{Eq:SN}
}
%----------------------------------------------------------------------------------------------------%
Note that the key inputs to determine SNR are $\sigma(D)$ and $\pdf(D)$. 

%<><><><><><><><><><><><><><><><><><><><><><><><><><><><><><><><><><><><><><><><><><><><><><><><><><>%
\begin{figure}[t]
\bc
\ifdefined\compile
\includegraphics[width=8.5cm,clip,height=6.5cm]{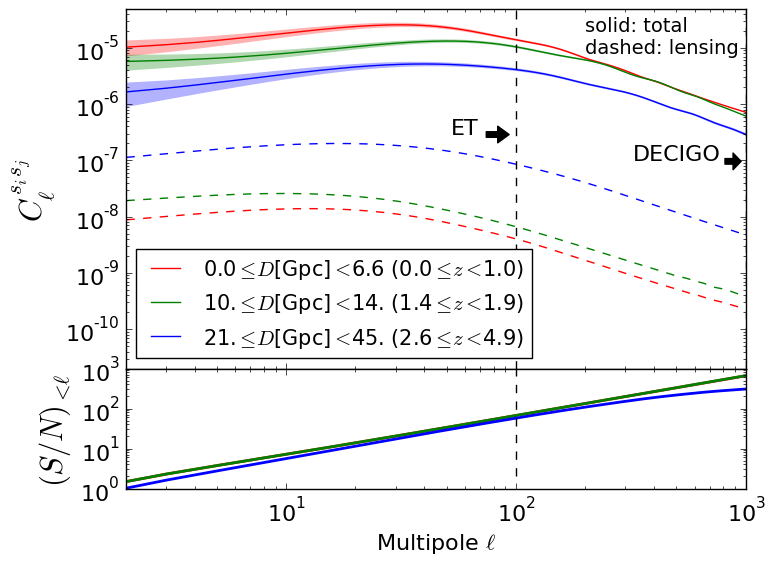} % = 135 words
\fi
\caption{
Angular auto-power spectra with expected $1\sigma$ statistical errors (top) and the cumulative SNR of 
the LSS-induced anisotropies (bottom) with $\Delta\l=1$. Assuming the three-year observation with the 
ET, we divide measured luminosity distances into five contiguous bins with equal number of binary 
sources. The angular power spectrum and SNR is then computed, and plotted as a function of the 
maximum multipole, $\l_{\rm max}$. The dashed lines show the lensing contributions. The vertical axis 
at $\l=100$ corresponds to the angular resolution of the ET while the resolution of DECIGO is 
approximately $\l=1000$. 
}
\label{Fig:SNR-ET}
\ec
\end{figure}
%<><><><><><><><><><><><><><><><><><><><><><><><><><><><><><><><><><><><><><><><><><><><><><><><><><>%

Figure~\ref{Fig:SNR-ET} shows the angular power spectrum and SNR of the LSS-induced anisotropies 
$\hmag_i$ at each luminosity distance bin, plotted against maximum multipoles used in the analysis. 
Here, we divide measured luminosity distances into five bins with equal number of binary sources, 
among which the results at representative three bins are particularly shown. 
The power spectra $C_\l^{s_is_i}$ are computed with the CMB Boltzmann code {\tt CAMB} \cite{CAMB}, 
assuming the flat Lambda-CDM model with fiducial cosmological parameters consistent 
with the seven-year WMAP results \cite{Komatsu:2010fb},
using {\tt Halofit} for computing the non-linear matter power spectrum \cite{Smith:2003,Takahashi:2012}.
The number distribution of binaries is computed using the fitting formula for the NS-NS merger rate 
in Ref.~\cite{Cutler:2005qq} normalized by the current merger rate, 
$\dot{n}_0=10^{-6}$ Mpc$^{-3}$yr$^{-1}$ \cite{LIGO:2010cfa}, 
which is an intermediate value among several predictions. The power spectrum of the number density 
fluctuation is computed using a linear bias model with a redshift dependence, parametrized by 
$b(z)=b_0+b_z/\mC{D}$, where $\mC{D}$ is the growth function \cite{Yamamoto:2006,Namikawa:2011}. 
The fiducial values of the bias parameters are $b_0=b_z=1$. The resultant SNRs in 
Fig.~\ref{Fig:SNR-ET} are increased monotonically up to $\l>1000$ and they look quite similar. 
With enough angular resolution of GW detectors, 
the detection of the LSS-induced anisotropies is highly anticipated out to rather distant sources. 

To better resolve the source position, however, we at least need three ET-like detectors at different sites. 
This situation may be optimistic, but has been assumed in the previous studies using EM counterparts of GW sources. 
Indeed, the determination of the source position is indispensable for the cosmology 
with standard sirens, and we just follow the assumption of three ET-like detectors at the same sites 
as aLIGO and VIRGO. 
Then, based on Ref.~\cite{Sidery2014PRD}, the angular resolutions for GW sources at $z=1$, $2$, and $3$ 
are estimated to be $\sim 2.0$, $3.2$, and $3.3 \,{\rm deg}$, respectively. 
The corresponding maximum multipoles become $\l_{\rm max}\sim90$, $57$, and $54$. 
Thus, a network of ET-like detectors is sufficient to provide an opportunity to constrain cosmology 
even at higher redshifts. This is rather contrasted to the previous study using redshift information, 
since identification of EM counterparts becomes harder as increasing the redshift. 
Since the SNR is limited by the cosmic variance, the estimated value of SNR at $\l_{\rm max}\sim100$ 
almost remains the same, irrespective of the uncertainties in the overall merger rate 
and $\sigma(D)$.
Note that for the space-based GW detector like DECIGO, the angular resolution would be much better 
even at $z>2$, and down to $\sim 0.027$ deg$^2$ ($\l_{\rm max}\sim1000$) \cite{Cutler:2009qv}. 

%////////////////////////////////////////////////////////////////////////////////////////////////////%
{\it Cosmological implications ---}
A statistically significant detection of the LSS-induced anisotropies has a potential to tightly 
constrain cosmology. Since the signal is expected to be statistically high even at distant luminosity distance bins 
(i.e., high redshift), the LSS-induced anisotropies would be powerful to constrain 
the early cosmic evolution. One interesting application may be the primordial non-Gaussianity, 
which can be imprinted on the large-scale clustering at higher redshifts through the scale-dependent 
galaxy bias (e.g., Refs.~\cite{Dalal:2008,Slosar:2008}). The leading-order effect of the primordial non-Gaussianity is 
characterized by the parameter $f_{\rm NL}$ \cite{Komatsu:2001}, and by looking at LSS-induced anisotropies 
at lower multipoles, a strong constraint on $f_{\rm NL}$ will be obtained. Taking a full covariance 
matrix of the LSS-induced signals $\hmag_i$ up to $\l_{\rm max}=100$, 
we estimate the expected constraints on the $f_{\rm NL}$ based on the Fisher matrix analysis. 
To be specific, in the Fisher matrix, we consider the five luminosity-distance bins as demonstrated 
in Fig.~\ref{Fig:SNR-ET}. We only marginalize three parameters $f_{\rm NL}$, $b_0$, and $b_z$, and 
compute the expected 1$\sigma$ errors. Note that degeneracy between $f_{\rm NL}$ and other cosmological 
parameters are very weak and we consider only these parameters \cite{Namikawa:2011}. 
The Fisher matrix analysis reveals that the marginalized error $\sigma(f_{\rm NL})=0.54$. 
This constraint is roughly comparable to or even better than those obtained from the future LSS surveys \cite{Carbone:2008}.
Discarding the two-distant bins still leads to a severe constraint, $\sigma(f_{\rm NL})=1.2$, indicating 
that primordial non-Gaussianity will be robustly constrained irrespective of a large 
uncertainty in the binary distribution.

The LSS-induced anisotropies would be also useful for cross-correlation studies with other cosmological probes. 
The interesting counterpart of the cross correlation would be the gravitational lensing of 
the cosmic microwave background (CMB) and weak lensing signals from galaxy surveys. 
Among the two contributions in the LSS-induced anisotropies of standard sirens [see Eq.~\eqref{Eq:est-theory}],
the former cross correlation picks up the clustering term in the luminosity distance anisotropies, 
and it provides a tomographic view of the galaxy clustering at higher redshifts. We find that 
the SNR at the most distant bin would be $31$ and $43$ combined with the Planck \cite{P15} and 
CMB Stage-IV \cite{Abazajian:2013oma}, respectively. 
On the other hand, the latter cross correlation enables us to extract 
the pure lensing signals in the luminosity distance anisotropies, 
since the clustering term in Eq.~\eqref{Eq:est-theory}
is mostly uncorrelated between different redshifts. 
With the future lensing measurement by Euclid \cite{Euclid}, the SNR is found to be 16. 

%////////////////////////////////////////////////////////////////////////////////////////////////////%
{\it Discussion.---}
So far, we have considered the NS binaries as the representative GW standard sirens, but the proposed 
method can be also applied to other GW sources which are not necessarily accompanied by EM counterparts. 
Examples are black-hole (BH) binaries and NS-BH binaries \cite{LIGO:2010cfa}. Combining these binaries 
with NS binaries would certainly increase the SNR of the LSS-induced anisotropies, further improving 
constraining power on cosmology. 

Potential systematics to the LSS-induced anisotropies may come from the luminosity distance estimation 
to each GW source. In our analysis, the luminosity distance error was estimated based on the averaged 
GW waveform over the inclination angle of NS binary, assuming the isotropic antenna pattern averaged 
over the sky. The antenna pattern of a GW detector is, however, anisotropic, and needs to be properly 
taken into account in practical data analysis. Further, the nonvanishing inclination of the NS binary 
is known to produce a strong parameter degeneracy with the luminosity distance \cite{Nissanke:2009kt}. 
This degeneracy can increase the measurement uncertainty in the luminosity distance especially for 
low SNR GW sources, potentially leading to a biased estimation of the anisotropic signal. Following 
Ref.~\cite{Nissanke:2009kt}, we estimate the increase of the uncertainty in the luminosity distance 
due to the degeneracy, and find that the uncertainty increases at most by a factor of 2. Since the 
detection significance of the LSS-induced anisotropies is almost determined by the cosmic variance, 
the degeneracy has negligible impact on our results.

%////////////////////////////////////////////////////////////////////////////////////////////////////%
{\it Summary.---}
We proposed a novel method to probe cosmology from the GW standard sirens without redshift information. 
A key observable is the anisotropies in the number distribution and luminosity distances to each GW source, 
arising from the clustering and lensing effect of the LSS. Based on a simple estimator given at 
Eq.~(\ref{Eq:est-mag}), feasibility to detect the LSS-induced anisotropies has been discussed, and we 
found that, via a network of ET-like detectors, once GW observations will be well established in the future, 
\bi
\item The anisotropies at very high redshift ($z>2$) can be detected at high statistical significance 
(SNR$\gtrsim 100$)  
\item Constraining power on the primordial non-Gaussianity is roughly comparable to or even better 
than those from the future LSS surveys, 
\item Cross correlation with other cosmological probes such as CMB and galaxy weak lensing would be 
also detectable at high significance by combining future CMB experiments and galaxy surveys observing 
at the same epoch. 
\ei
Albeit simple, the present method offers a direct way to probe cosmology only from the GW measurement, 
and this would provide new insight into the formation and evolution of large-scale structure, 
definitely complementary to the EM observations. 

%////////////////////////////////////////////////////////////////////////////////////////////////////%
{\it Note added.---}
After our paper was accepted, the first detection of GW from binary BH has been reported \cite{GW150914}, 
and this enlarges the future prospect for measuring anisotropic signals. 
First, the detection suggests a rather higher merger rate for binary BHs, $2$--$400$ Gpc$^{-3}$ yr$^{-1}$, 
indicating that even the second-generation GW detectors have a potential to detect the anisotropies of 
binary BH sources. A high merger rate also suggests that a large number of binary BH will be observed 
at milli-Hz band, and with the eLISA, most of the extra-galactic sources will be spatially resolved 
\cite{Sesana:2016} (see \cite{Seto:2016} for Galactic BH binaries). 
This implies that in an optimistic case, a single ET-like detector is sufficient to detect 
the anisotropic signals in combination with the eLISA measurements. 

% Acknowledgments %
\ifdefined\compile
\begin{acknowledgments}
This work is supported in part by JSPS Postdoctoral Fellowships for Research Abroad No.~26-142 (T.N.)., 
No.~25-180 (A.N.). This work is in part supported by MEXT KAKENHI (15H05889 for AT).
\end{acknowledgments}

% References %
\bibliographystyle{apsrev}
\bibliography{cite}

\begin{thebibliography}{43}
\expandafter\ifx\csname natexlab\endcsname\relax\def\natexlab#1{#1}\fi
\expandafter\ifx\csname bibnamefont\endcsname\relax
  \def\bibnamefont#1{#1}\fi
\expandafter\ifx\csname bibfnamefont\endcsname\relax
  \def\bibfnamefont#1{#1}\fi
\expandafter\ifx\csname citenamefont\endcsname\relax
  \def\citenamefont#1{#1}\fi
\expandafter\ifx\csname url\endcsname\relax
  \def\url#1{\texttt{#1}}\fi
\expandafter\ifx\csname urlprefix\endcsname\relax\def\urlprefix{URL }\fi
\providecommand{\bibinfo}[2]{#2}
\providecommand{\eprint}[2][]{\url{#2}}

\bibitem[{\citenamefont{{LIGO Scientific Collaboration and Virgo
  Collaboration}}(2016)}]{GW150914}
\bibinfo{author}{\bibnamefont{{LIGO Scientific Collaboration and Virgo
  Collaboration}}}, \bibinfo{journal}{\prl} \textbf{\bibinfo{volume}{116}},
  \bibinfo{pages}{061102} (\bibinfo{year}{2016}).

\bibitem[{\citenamefont{Evans}(2014)}]{Evans2014GRG}
\bibinfo{author}{\bibfnamefont{M.}~\bibnamefont{Evans}}, \bibinfo{journal}{Gen.
  Relativ. Gravit.} \textbf{\bibinfo{volume}{46}}, \bibinfo{pages}{1778}
  (\bibinfo{year}{2014}).

\bibitem[{\citenamefont{Punturo et~al.}(2010)}]{Punturo:2010}
\bibinfo{author}{\bibfnamefont{M.}~\bibnamefont{Punturo}} \bibnamefont{et~al.},
  \bibinfo{journal}{Class. Quant. Grav.} \textbf{\bibinfo{volume}{27}},
  \bibinfo{pages}{194002} (\bibinfo{year}{2010}).

\bibitem[{\citenamefont{Dwyer et~al.}(2015)\citenamefont{Dwyer, Sigg, Ballmer,
  Barsotti, Mavalvala, and Evans}}]{Dwyer2015PRD}
\bibinfo{author}{\bibfnamefont{S.}~\bibnamefont{Dwyer}},
  \bibinfo{author}{\bibfnamefont{D.}~\bibnamefont{Sigg}},
  \bibinfo{author}{\bibfnamefont{S.~W.} \bibnamefont{Ballmer}},
  \bibinfo{author}{\bibfnamefont{L.}~\bibnamefont{Barsotti}},
  \bibinfo{author}{\bibfnamefont{N.}~\bibnamefont{Mavalvala}},
  \bibnamefont{and} \bibinfo{author}{\bibfnamefont{M.}~\bibnamefont{Evans}},
  \bibinfo{journal}{\prd} \textbf{\bibinfo{volume}{91}},
  \bibinfo{pages}{082001} (\bibinfo{year}{2015}).

\bibitem[{\citenamefont{Seoane et~al.}(2013)}]{Seoane:2013qna}
\bibinfo{author}{\bibfnamefont{P.~A.} \bibnamefont{Seoane}}
  \bibnamefont{et~al.} (\bibinfo{collaboration}{eLISA Collaboration})
  (\bibinfo{year}{2013}), \eprint{arXiv: 1305.5720}.

\bibitem[{\citenamefont{Sato et~al.}(2009)}]{Sato:2009}
\bibinfo{author}{\bibfnamefont{S.}~\bibnamefont{Sato}} \bibnamefont{et~al.},
  \bibinfo{journal}{J. of Phys. Conf. Series} \textbf{\bibinfo{volume}{154}},
  \bibinfo{pages}{012040} (\bibinfo{year}{2009}).

\bibitem[{\citenamefont{Schutz}(1986)}]{Schutz1986Nature}
\bibinfo{author}{\bibfnamefont{B.~F.} \bibnamefont{Schutz}},
  \bibinfo{journal}{\nat} \textbf{\bibinfo{volume}{323}}, \bibinfo{pages}{310}
  (\bibinfo{year}{1986}).

\bibitem[{\citenamefont{Petiteau et~al.}(2011)\citenamefont{Petiteau, Babak,
  and Sesana}}]{Petiteau2011ApJ}
\bibinfo{author}{\bibfnamefont{A.}~\bibnamefont{Petiteau}},
  \bibinfo{author}{\bibfnamefont{S.}~\bibnamefont{Babak}}, \bibnamefont{and}
  \bibinfo{author}{\bibfnamefont{A.}~\bibnamefont{Sesana}},
  \bibinfo{journal}{\apj} \textbf{\bibinfo{volume}{732}}, \bibinfo{pages}{82}
  (\bibinfo{year}{2011}).

\bibitem[{\citenamefont{Cutler and Holz}(2009)}]{Cutler:2009qv}
\bibinfo{author}{\bibfnamefont{C.}~\bibnamefont{Cutler}} \bibnamefont{and}
  \bibinfo{author}{\bibfnamefont{D.~E.} \bibnamefont{Holz}},
  \bibinfo{journal}{\prd} \textbf{\bibinfo{volume}{80}},
  \bibinfo{pages}{104009} (\bibinfo{year}{2009}).

\bibitem[{\citenamefont{Nishizawa et~al.}(2011)\citenamefont{Nishizawa, Taruya,
  and Saito}}]{Nishizawa:2010xx}
\bibinfo{author}{\bibfnamefont{A.}~\bibnamefont{Nishizawa}},
  \bibinfo{author}{\bibfnamefont{A.}~\bibnamefont{Taruya}}, \bibnamefont{and}
  \bibinfo{author}{\bibfnamefont{S.}~\bibnamefont{Saito}},
  \bibinfo{journal}{\prd} \textbf{\bibinfo{volume}{83}},
  \bibinfo{pages}{084045} (\bibinfo{year}{2011}).

\bibitem[{\citenamefont{Sathyaprakash et~al.}(2010)\citenamefont{Sathyaprakash,
  Schutz, and Van Den~Broeck}}]{Sathyaprakash:2009xt}
\bibinfo{author}{\bibfnamefont{B.}~\bibnamefont{Sathyaprakash}},
  \bibinfo{author}{\bibfnamefont{B.}~\bibnamefont{Schutz}}, \bibnamefont{and}
  \bibinfo{author}{\bibfnamefont{C.}~\bibnamefont{Van Den~Broeck}},
  \bibinfo{journal}{Class.Quant.Grav.} \textbf{\bibinfo{volume}{27}},
  \bibinfo{pages}{215006} (\bibinfo{year}{2010}).

\bibitem[{\citenamefont{Taylor and Gair}(2012)}]{Taylor:2012db}
\bibinfo{author}{\bibfnamefont{S.~R.} \bibnamefont{Taylor}} \bibnamefont{and}
  \bibinfo{author}{\bibfnamefont{J.~R.} \bibnamefont{Gair}},
  \bibinfo{journal}{\prd} \textbf{\bibinfo{volume}{86}},
  \bibinfo{pages}{023502} (\bibinfo{year}{2012}).

\bibitem[{\citenamefont{Hirata et~al.}(2010)\citenamefont{Hirata, Holz, and
  Cutler}}]{Hirata:2010}
\bibinfo{author}{\bibfnamefont{C.~M.} \bibnamefont{Hirata}},
  \bibinfo{author}{\bibfnamefont{D.~E.} \bibnamefont{Holz}}, \bibnamefont{and}
  \bibinfo{author}{\bibfnamefont{C.}~\bibnamefont{Cutler}},
  \bibinfo{journal}{\prd} \textbf{\bibinfo{volume}{81}},
  \bibinfo{pages}{124046} (\bibinfo{year}{2010}).

\bibitem[{\citenamefont{Shang and Haiman}(2011)}]{Shang2011MNRAS}
\bibinfo{author}{\bibfnamefont{C.}~\bibnamefont{Shang}} \bibnamefont{and}
  \bibinfo{author}{\bibfnamefont{Z.}~\bibnamefont{Haiman}},
  \bibinfo{journal}{Mon. Not. Roy. Astron. Soc.}
  \textbf{\bibinfo{volume}{411}}, \bibinfo{pages}{9} (\bibinfo{year}{2011}).

\bibitem[{\citenamefont{Camera and Nishizawa}(2013)}]{Camera:2013xfa}
\bibinfo{author}{\bibfnamefont{S.}~\bibnamefont{Camera}} \bibnamefont{and}
  \bibinfo{author}{\bibfnamefont{A.}~\bibnamefont{Nishizawa}},
  \bibinfo{journal}{\prl} \textbf{\bibinfo{volume}{110}},
  \bibinfo{pages}{151103} (\bibinfo{year}{2013}).

\bibitem[{\citenamefont{Cooray et~al.}(2006)\citenamefont{Cooray, Holz, and
  Huterer}}]{Cooray:2006}
\bibinfo{author}{\bibfnamefont{A.}~\bibnamefont{Cooray}},
  \bibinfo{author}{\bibfnamefont{D.}~\bibnamefont{Holz}}, \bibnamefont{and}
  \bibinfo{author}{\bibfnamefont{D.}~\bibnamefont{Huterer}},
  \bibinfo{journal}{\apj} \textbf{\bibinfo{volume}{637}}, \bibinfo{pages}{L77}
  (\bibinfo{year}{2006}).

\bibitem[{\citenamefont{McQuinn and White}(2013)}]{McQuinn:2013}
\bibinfo{author}{\bibfnamefont{M.}~\bibnamefont{McQuinn}} \bibnamefont{and}
  \bibinfo{author}{\bibfnamefont{M.}~\bibnamefont{White}},
  \bibinfo{journal}{Mon. Not. Roy. Astron. Soc.}
  \textbf{\bibinfo{volume}{433}}, \bibinfo{pages}{2857} (\bibinfo{year}{2013}).

\bibitem[{\citenamefont{Nissanke et~al.}(2013)\citenamefont{Nissanke, Kasliwal,
  and Georgieva}}]{Nissanke:2013}
\bibinfo{author}{\bibfnamefont{S.}~\bibnamefont{Nissanke}},
  \bibinfo{author}{\bibfnamefont{M.}~\bibnamefont{Kasliwal}}, \bibnamefont{and}
  \bibinfo{author}{\bibfnamefont{A.}~\bibnamefont{Georgieva}},
  \bibinfo{journal}{\apj} \textbf{\bibinfo{volume}{767}}, \bibinfo{pages}{124}
  (\bibinfo{year}{2013}).

\bibitem[{\citenamefont{Nishizawa et~al.}(2012)\citenamefont{Nishizawa, Yagi,
  Taruya, and Tanaka}}]{Nishizawa:2011eq}
\bibinfo{author}{\bibfnamefont{A.}~\bibnamefont{Nishizawa}},
  \bibinfo{author}{\bibfnamefont{K.}~\bibnamefont{Yagi}},
  \bibinfo{author}{\bibfnamefont{A.}~\bibnamefont{Taruya}}, \bibnamefont{and}
  \bibinfo{author}{\bibfnamefont{T.}~\bibnamefont{Tanaka}},
  \bibinfo{journal}{\prd} \textbf{\bibinfo{volume}{85}},
  \bibinfo{pages}{044047} (\bibinfo{year}{2012}).

\bibitem[{\citenamefont{MacLeod and Hogan}(2008)}]{MacLeod2008PRD}
\bibinfo{author}{\bibfnamefont{C.~L.} \bibnamefont{MacLeod}} \bibnamefont{and}
  \bibinfo{author}{\bibfnamefont{C.~J.} \bibnamefont{Hogan}},
  \bibinfo{journal}{\prd} \textbf{\bibinfo{volume}{77}},
  \bibinfo{pages}{043512} (\bibinfo{year}{2008}).

\bibitem[{\citenamefont{Messenger and Read}(2012)}]{Messenger:2011gi}
\bibinfo{author}{\bibfnamefont{C.}~\bibnamefont{Messenger}} \bibnamefont{and}
  \bibinfo{author}{\bibfnamefont{J.}~\bibnamefont{Read}},
  \bibinfo{journal}{\prl} \textbf{\bibinfo{volume}{108}},
  \bibinfo{pages}{091101} (\bibinfo{year}{2012}).

\bibitem[{\citenamefont{Taylor et~al.}(2012)\citenamefont{Taylor, Gair, and
  Mandel}}]{Taylor:2011fs}
\bibinfo{author}{\bibfnamefont{S.~R.} \bibnamefont{Taylor}},
  \bibinfo{author}{\bibfnamefont{J.~R.} \bibnamefont{Gair}}, \bibnamefont{and}
  \bibinfo{author}{\bibfnamefont{I.}~\bibnamefont{Mandel}},
  \bibinfo{journal}{\prd} \textbf{\bibinfo{volume}{85}},
  \bibinfo{pages}{023535} (\bibinfo{year}{2012}).

\bibitem[{\citenamefont{Futamase and Sasaki}(1989)}]{FutamaseSasaki:1989}
\bibinfo{author}{\bibfnamefont{T.}~\bibnamefont{Futamase}} \bibnamefont{and}
  \bibinfo{author}{\bibfnamefont{M.}~\bibnamefont{Sasaki}},
  \bibinfo{journal}{Phys. Rev.} \textbf{\bibinfo{volume}{D40}},
  \bibinfo{pages}{2502} (\bibinfo{year}{1989}).

\bibitem[{\citenamefont{Regimbau et~al.}(2012)}]{Regimbau:2012ir}
\bibinfo{author}{\bibfnamefont{T.}~\bibnamefont{Regimbau}}
  \bibnamefont{et~al.}, \bibinfo{journal}{\prd} \textbf{\bibinfo{volume}{86}},
  \bibinfo{pages}{122001} (\bibinfo{year}{2012}).

\bibitem[{\citenamefont{Lewis et~al.}(2000)\citenamefont{Lewis, Challinor, and
  Lasenby}}]{CAMB}
\bibinfo{author}{\bibfnamefont{A.}~\bibnamefont{Lewis}},
  \bibinfo{author}{\bibfnamefont{A.}~\bibnamefont{Challinor}},
  \bibnamefont{and} \bibinfo{author}{\bibfnamefont{A.}~\bibnamefont{Lasenby}},
  \bibinfo{journal}{\apj} \textbf{\bibinfo{volume}{538}}, \bibinfo{pages}{473}
  (\bibinfo{year}{2000}).

\bibitem[{\citenamefont{Komatsu et~al.}(2011)}]{Komatsu:2010fb}
\bibinfo{author}{\bibfnamefont{E.}~\bibnamefont{Komatsu}} \bibnamefont{et~al.},
  \bibinfo{journal}{\apj} \textbf{\bibinfo{volume}{192}}, \bibinfo{pages}{18}
  (\bibinfo{year}{2011}).

\bibitem[{\citenamefont{Smith et~al.}(2003)}]{Smith:2003}
\bibinfo{author}{\bibfnamefont{R.~E.} \bibnamefont{Smith}}
  \bibnamefont{et~al.}, \bibinfo{journal}{Mon. Not. Roy. Astron. Soc.}
  \textbf{\bibinfo{volume}{341}}, \bibinfo{pages}{1311} (\bibinfo{year}{2003}).

\bibitem[{\citenamefont{Takahashi et~al.}(2012)}]{Takahashi:2012}
\bibinfo{author}{\bibfnamefont{R.}~\bibnamefont{Takahashi}}
  \bibnamefont{et~al.}, \bibinfo{journal}{\apj} \textbf{\bibinfo{volume}{761}},
  \bibinfo{pages}{152} (\bibinfo{year}{2012}).

\bibitem[{\citenamefont{Cutler and Harms}(2006)}]{Cutler:2005qq}
\bibinfo{author}{\bibfnamefont{C.}~\bibnamefont{Cutler}} \bibnamefont{and}
  \bibinfo{author}{\bibfnamefont{J.}~\bibnamefont{Harms}},
  \bibinfo{journal}{\prd} \textbf{\bibinfo{volume}{73}},
  \bibinfo{pages}{042001} (\bibinfo{year}{2006}).

\bibitem[{\citenamefont{Abadie et~al.}(2010)}]{LIGO:2010cfa}
\bibinfo{author}{\bibfnamefont{J.}~\bibnamefont{Abadie}} \bibnamefont{et~al.}
  (\bibinfo{collaboration}{LIGO Scientific}), \bibinfo{journal}{Class. Quant.
  Grav.} \textbf{\bibinfo{volume}{27}}, \bibinfo{pages}{173001}
  (\bibinfo{year}{2010}).

\bibitem[{\citenamefont{Yamamoto et~al.}(2006)}]{Yamamoto:2006}
\bibinfo{author}{\bibfnamefont{K.}~\bibnamefont{Yamamoto}}
  \bibnamefont{et~al.}, \bibinfo{journal}{\prd} \textbf{\bibinfo{volume}{74}},
  \bibinfo{pages}{063525} (\bibinfo{year}{2006}).

\bibitem[{\citenamefont{{Namikawa} et~al.}(2011)\citenamefont{{Namikawa},
  {Okamura}, and {Taruya}}}]{Namikawa:2011}
\bibinfo{author}{\bibfnamefont{T.}~\bibnamefont{{Namikawa}}},
  \bibinfo{author}{\bibfnamefont{T.}~\bibnamefont{{Okamura}}},
  \bibnamefont{and} \bibinfo{author}{\bibfnamefont{A.}~\bibnamefont{{Taruya}}},
  \bibinfo{journal}{\prd} \textbf{\bibinfo{volume}{83}},
  \bibinfo{pages}{123514} (\bibinfo{year}{2011}).

\bibitem[{\citenamefont{Sidery et~al.}(2014)\citenamefont{Sidery, Aylott,
  Christensen, Farr, Farr, Feroz, Gair, Grover, Graff, Hanna
  et~al.}}]{Sidery2014PRD}
\bibinfo{author}{\bibfnamefont{T.}~\bibnamefont{Sidery}},
  \bibinfo{author}{\bibfnamefont{B.}~\bibnamefont{Aylott}},
  \bibinfo{author}{\bibfnamefont{N.}~\bibnamefont{Christensen}},
  \bibinfo{author}{\bibfnamefont{B.}~\bibnamefont{Farr}},
  \bibinfo{author}{\bibfnamefont{W.}~\bibnamefont{Farr}},
  \bibinfo{author}{\bibfnamefont{F.}~\bibnamefont{Feroz}},
  \bibinfo{author}{\bibfnamefont{J.}~\bibnamefont{Gair}},
  \bibinfo{author}{\bibfnamefont{K.}~\bibnamefont{Grover}},
  \bibinfo{author}{\bibfnamefont{P.}~\bibnamefont{Graff}},
  \bibinfo{author}{\bibfnamefont{C.}~\bibnamefont{Hanna}},
  \bibnamefont{et~al.}, \bibinfo{journal}{\prd} \textbf{\bibinfo{volume}{89}},
  \bibinfo{pages}{084060} (\bibinfo{year}{2014}).

\bibitem[{\citenamefont{Dalal et~al.}(2008)}]{Dalal:2008}
\bibinfo{author}{\bibfnamefont{N.}~\bibnamefont{Dalal}} \bibnamefont{et~al.},
  \bibinfo{journal}{\prd} \textbf{\bibinfo{volume}{77}},
  \bibinfo{pages}{123514} (\bibinfo{year}{2008}).

\bibitem[{\citenamefont{Slosar et~al.}(2008)}]{Slosar:2008}
\bibinfo{author}{\bibfnamefont{A.}~\bibnamefont{Slosar}} \bibnamefont{et~al.},
  \bibinfo{journal}{J. Cosmol. Astropart. Phys.} \textbf{\bibinfo{volume}{08}},
  \bibinfo{pages}{031} (\bibinfo{year}{2008}).

\bibitem[{\citenamefont{Komatsu and Spergel}(2001)}]{Komatsu:2001}
\bibinfo{author}{\bibfnamefont{E.}~\bibnamefont{Komatsu}} \bibnamefont{and}
  \bibinfo{author}{\bibfnamefont{D.~N.} \bibnamefont{Spergel}},
  \bibinfo{journal}{\prd} \textbf{\bibinfo{volume}{63}},
  \bibinfo{pages}{063002} (\bibinfo{year}{2001}).

\bibitem[{\citenamefont{Carbone et~al.}(2008)\citenamefont{Carbone, Verde, and
  Matarrese}}]{Carbone:2008}
\bibinfo{author}{\bibfnamefont{C.}~\bibnamefont{Carbone}},
  \bibinfo{author}{\bibfnamefont{L.}~\bibnamefont{Verde}}, \bibnamefont{and}
  \bibinfo{author}{\bibfnamefont{S.}~\bibnamefont{Matarrese}},
  \bibinfo{journal}{\apj} \textbf{\bibinfo{volume}{684}}, \bibinfo{pages}{L1}
  (\bibinfo{year}{2008}).

\bibitem[{\citenamefont{{Planck Collaboration}}(2015)}]{P15}
\bibinfo{author}{\bibnamefont{{Planck Collaboration}}} (\bibinfo{year}{2015}),
  \eprint{arXiv:1502.01591}.

\bibitem[{\citenamefont{Abazajian et~al.}(2015)}]{Abazajian:2013oma}
\bibinfo{author}{\bibfnamefont{K.}~\bibnamefont{Abazajian}}
  \bibnamefont{et~al.}, \bibinfo{journal}{Astropart. Phys.}
  \textbf{\bibinfo{volume}{63}}, \bibinfo{pages}{66} (\bibinfo{year}{2015}).

\bibitem[{\citenamefont{{The Euclid Theory Working Group}}(2013)}]{Euclid}
\bibinfo{author}{\bibnamefont{{The Euclid Theory Working Group}}},
  \bibinfo{journal}{Living Rev. Relativity} \textbf{\bibinfo{volume}{16}},
  \bibinfo{pages}{6} (\bibinfo{year}{2013}).

\bibitem[{\citenamefont{Nissanke et~al.}(2010)\citenamefont{Nissanke, Holz,
  Hughes, Dalal, and Sievers}}]{Nissanke:2009kt}
\bibinfo{author}{\bibfnamefont{S.}~\bibnamefont{Nissanke}},
  \bibinfo{author}{\bibfnamefont{D.~E.} \bibnamefont{Holz}},
  \bibinfo{author}{\bibfnamefont{S.~A.} \bibnamefont{Hughes}},
  \bibinfo{author}{\bibfnamefont{N.}~\bibnamefont{Dalal}}, \bibnamefont{and}
  \bibinfo{author}{\bibfnamefont{J.~L.} \bibnamefont{Sievers}},
  \bibinfo{journal}{\apj} \textbf{\bibinfo{volume}{725}}, \bibinfo{pages}{496}
  (\bibinfo{year}{2010}).

\bibitem[{\citenamefont{Sesana}(2016)}]{Sesana:2016}
\bibinfo{author}{\bibfnamefont{A.}~\bibnamefont{Sesana}}
  (\bibinfo{year}{2016}), \eprint{arXiv:1602.06951}.

\bibitem[{\citenamefont{Seto}(2016)}]{Seto:2016}
\bibinfo{author}{\bibfnamefont{N.}~\bibnamefont{Seto}} (\bibinfo{year}{2016}),
  \eprint{arXiv:1602.04715}.

\end{thebibliography}
\fi

\end{document}